\documentclass[aps,prb,twocolumn,showpacs]{revtex4}
\usepackage{graphicx}

\begin{document}

\title{BaFe$_2$Se$_2$O as an Iron-Based Mott Insulator with Antiferromagnetic Order}
\author{Fei Han$^{1,2}$, Xiangang Wan$^{1}$, Bing Shen$^{1,2}$, and Hai-Hu
Wen$^{1}$}
\email{hhwen@nju.edu.cn}
\affiliation{$^{1}$Center for Superconducting Physics and Materials, National Laboratory
of Solid State Microstructures and Department of Physics, Nanjing
University, Nanjing 210093, China}
\affiliation{$^{2}$National Laboratory for Superconductivity, Institute of Physics and
Beijing National Laboratory for Condensed Matter Physics, Chinese Academy of
Sciences, Beijing 100190, China}

\begin{abstract}
A iron-based compound with a quasi-two-dimensional array of FeSe$_{3}$O
tetrahedra and an orthorhombic structure, namely
BaFe$_{2}$Se$_{2}$O, has been successfully fabricated.
Experimental results show that this compound is an insulator and
has an antiferromagnetic (AF) transition at 240 K. Band structure
calculation reveals the narrowing of Fe 3$d$ bands near the Fermi
energy, which leads to the localization of magnetism and the Mott
insulating behavior. The large distances between the Fe atoms perhaps are responsible for the characters. Linear response calculation further indicates
a strong in-plane AF exchange $J$, this can account
for the enhanced magnetic susceptibility (which has a maximum at
about 450 K) above the Neel temperature.
\end{abstract}

\maketitle

\subsection{Introduction}

Discovery of the copper-based superconductors\cite{HighTc} with high
transition temperatures ($T_c$) has attracted great attention and
triggered extensive research in the fields of condensed matter
physics and material sciences. High-$T_c$ superconductivity in the
cuprates occurs with either hole or electron doping in the
antiferromagnetic (AF) Mott insulating parent compounds. The
Mottness clearly plays an important role here\cite{Anderson}. In the
past years, scientists have been trying to explore new high-$T_c$
superconductors in non-copper-based compounds. Until four years ago,
the discovery of superconductivity at 26 K in oxy-arsenide
LaFeAsO$_{1-x}$F$_{x}$\cite{Hosono} broke the uniqueness of cuprates
in giving rise to high-$T_c$ superconductivity. This is actually a
surprise since Fe atoms are normally strongly magnetic in materials
and usually inhibit superconductivity. Among the iron-based
superconductors with several different
structures\cite{Hosono,Rotter,ChuCW,WangXC,ChuCW2,WuMK,Cava,ZhuXY,Cava2},
it is widely perceived that the common Fe$X$ ($X$ = pnictide or
chalcogenide) layers are responsible for the superconductivity.
Meanwhile, for many FeAs-based superconductors, their parent phases
possess an AF order which seems to compete with the
superconductivity. To unravel the mystery of the origin of magnetism
and its relation with superconductivity is the most important issue.
Now there exist two major pictures, namely the electron pairing is
established through a weak coupling by exchanging the AF spin
fluctuations\cite{Mazin,Kuroki}, or the pairing is furnished by the
local strong pairing\cite{Kotliar,SiQM}. Related to this debate, it
has been argued whether the magnetism in the iron-based
superconductors is localized or itinerant in nature. It thus becomes
very important to find a iron-based compound with a formally
known knowledge of localized or itinerant magnetism. Recently,
exploration along this line was successful in finding the materials
of $Re_{2}$O$_{2}$Fe$_{2}$O(Se/S)$_{2}$\cite{ZhuJX,NiN} ($Re$ = rare earth element),
La$_{2}$Co$_{2}$Se$_{2}$O$_{3}$\cite{WangC}, La$_{2}$O$_{3}$Mn$_{2}$Se$_{2}$\cite{NiN2} and
Na$_{2}$Fe$_{2}$Se$_{2}$O\cite{ChenGF}. These compounds all contain
an expanded similar square-planar $M$ lattice in their
$M_{2}$OSe$_{2}$ ($M$ = Fe, Co and Mn) layers. Here we report the fabrication of an
iron-based compound BaFe$_{2}$Se$_{2}$O. From the structure point of view,
it looks like the existing materials
$Re_{2}$O$_{2}M_{2}$O(Se/S)$_{2}$ and Na$_{2}$Fe$_{2}$Se$_{2}$O.
However, a closer scrutiny finds the clear difference between them,
since the compound BaFe$_{2}$Se$_{2}$O contains a honeycomb-like
Fe sub-lattice instead of the square-planar Fe lattice. The compound
has an orthorhombic structure and is revealed to be a Mott insulator
with a localized antiferromagnetism.

\subsection{Experimental Details}

The single-crystalline sample was synthesized with the starting materials
BaO and FeSe in the ratio of 1:2 by using a flux method, similar to a
previous synthesis of K$_x$Fe$_{2-y}$Se$_2$\cite{ChenXL}. The resultant
crystals were black shining platelets with a typical dimension of 1.5 $%
\times $ 1.5 $\times$ 0.5 mm$^{3}$. To determine the compositional
ratio of atoms in the compound, scanning electron microscope and
energy-dispersive x-ray spectroscopy (EDX) measurements were carried
out on the crystals. Powder X-ray diffraction (XRD) pattern was
measured on the smashed and powdered crystals with $\theta-2\theta$
scan at room temperature. The XRD data were analyzed by a Rietveld
method with the software GSAS\cite{GSAS}. Resistivity measurement
was done using a four-probe technique on the physical property
measurement system (PPMS). Dc magnetic susceptibility was measured
on a superconducting quantum interference device (SQUID). Specific
heat was also measured using a relaxation technique on the PPMS.

\subsection{Experimental Data and Discussion}

\subsubsection{Crystal Structure}

The energy-dispersive x-ray spectroscopy (EDX) of the crystals is
shown in Fig. 1(a). One can see that the crystals are composed of
Ba, Fe, Se and O elements. The ratio of atoms is determined as
Ba:Fe:Se = 20.61:39.94:39.45 (close to 1:2:2) while the amounts of
O can not be quantified accurately. The EDX result offers us a
basis for seeking suitable reference compound to
do phase identification. We finalize our compound isostructural to $\beta $%
-BaFe$_{2}$S$_{3}$ (formerly reported as Ba$_{2}$Fe$_{4}$S$_{5}$\cite%
{Ba2Fe4S5,Ba2Fe4S52}). $\beta$-BaFe$_{2}$S$_{3}$ may be viewed as an
average due to twinning of a polymorph of $\alpha
$-BaFe$_{2}$S$_{3}$ which contains iron-deficient-FeAs-layers-like
Fe$_{2}$S$_{3}$ layers\cite{HongHY,Sefat,Petrovic}.
BaFe$_{2}$Se$_{2}$O can be considered as a derivative of
$\beta$-BaFe$_{2}$S$_{3}$ by replacing the S atoms at 2b sites with
O atoms and the S atoms at 4e sites with Se atoms. The powder x-ray
diffraction (XRD) pattern and Rietveld fit for the smashed and
powdered crystals is shown in Fig. 1(b). The data is well fitted
with an orthorhombic BaFe$_{2}$Se$_{2}$O phase which has the space
group of $Pmmn$. The unit cell is given in the inset of Fig. 1(b).
The starting parameters for the fit are taken from $\beta
$-BaFe$_{2}$S$_{3}$\cite{Ba2Fe4S52} and the GSAS program finally
finds the best fitting parameters. In Table I, the best fitting
structure parameters are listed. As shown in Fig. 1(c), in
BaFe$_{2}$Se$_{2}$O, the nearest neighbor atoms of each Fe atom,
namely the three Se atoms and one
O atom, surround the Fe atom and form a tetrahedra. The structure of BaFe$_{2}$Se$_{2}$O can also be viewed as an infinite quasi-two-dimensional array of FeSe$%
_{3}$O tetrahedra. Upper and lower Fe ladders along the $a$
direction and zig-zag Fe chains along the $b$ direction forms uneven
Fe layers. Obviously, there is not a strictly speaking square-planar
Fe lattice in this compound. It is interesting to notice that the
upper and lower Fe ladders form a non-coplanar honeycomb-like
structure, as shown in Fig. 1(d).

\begin{figure}
\includegraphics[width=8.5cm]{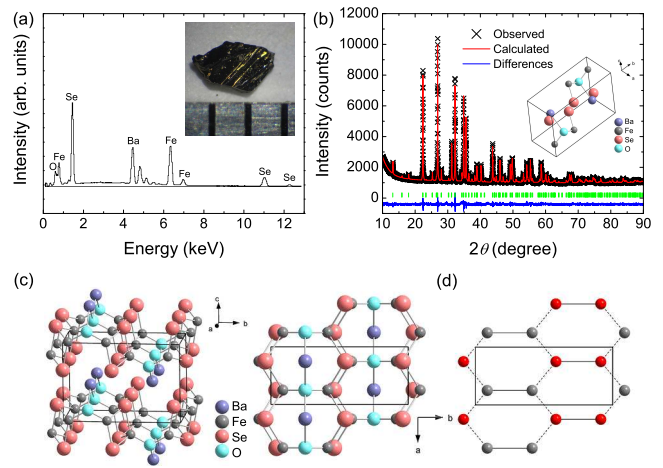}
\caption{(Color online) (a) Energy-dispersive x-ray spectroscopy (EDX) of
the crystal. The inset shows a crystal with the typical dimension of 1.5 $%
\times$ 1.5 $\times$ 0.5 mm$^{3}$. (b) Powder X-ray powder diffraction (XRD)
pattern and Rietveld fit for the smashed and powdered crystals. The inset
shows a unit cell of BaFe$_2$Se$_2$O. (c) Crystal Structure of BaFe$_2$Se$_2$%
O. (d) Honeycomb-like Fe sub-lattice in BaFe$_2$Se$_2$O, the gray
balls and the red balls all indicate Fe atoms, with the gray balls
forming the upper ladder and the red balls forming the lower ladder,
respectively.} \label{fig1}
\end{figure}

\begin{table}
\caption{Crystallographic data of BaFe$_{2}$Se$_{2}$O.}\tabcolsep 0pt
\vspace*{-24pt}
\par
\begin{center}
\def\temptablewidth{0.5\textwidth}
{\rule{\temptablewidth}{1pt}}
\begin{tabular*}{\temptablewidth}{@{\extracolsep{\fill}}ccccccc}
Space group & $Pmmn$ &  &  &  &  &  \\
$a$ ({\AA }), $b$ ({\AA }), $c$ ({\AA }) & 4.14028(5), 9.86716(14), 6.73381(8) &  &  &  &
&  \\
$V$({\AA }$^{3}$) & 275.095 &  &  &  &  &  \\
$Z$ & 2 &  &  &  &  &  \\
R$_p$, $w$R$_p$ & 2.41$\%$, 3.12$\%$ &  &  &  &  &  \\
Atomic parameters: &  &  &  &  &  &  \\
Ba & 2a ($\frac{1}{4}$, $\frac{1}{4}$, $z$) &  &  &  &  &  \\
& $z$=0.50850(18) &  &  &  &  &  \\
Fe & 4e ($\frac{1}{4}$, $y$, $z$) &  &  &  &  &  \\
& $y$=0.91347(22), $z$=0.87920(27) &  &  &  &  &  \\
Se & 4e ($\frac{1}{4}$, $y$, $z$) &  &  &  &  &  \\
& $y$=0.95840(15), $z$=0.24251(20) &  &  &  &  &  \\
O & 2b ($\frac{1}{4}$, $\frac{3}{4}$, $z$) &  &  &  &  &  \\
& $z$=0.7502(13) &  &  &  &  &  \\
Bonding lengths ({\AA }): &  &  &  &  &  &  \\
Ba-O & 2.706(6)$\times$2 &  &  &  &  &  \\
Fe-O & 1.832(4)$\times$1 &  &  &  &  &  \\
Fe-Se & 2.5604(15)$\times$2 &  &  &  &  &  \\
Fe-Se & 2.4863(23)$\times$1 &  &  &  &  &  \\
Fe-Fe & 3.1382(13)$\times$2 &  &  &  &  &  \\
Fe-Fe & 3.2260(12)$\times$1 &  &  &  &  &  \\
Bonding angles (degree): &  &  &  &  &  &  \\
Se-Fe-Se & 103.11(6)$\times$2 &  &  &  &  &  \\
Se-Fe-Se & 107.90(9)$\times$1 &  &  &  &  &  \\
Se-Fe-O & 106.44(14)$\times$2 &  &  &  &  &  \\
Se-Fe-O & 128.57(29)$\times$1 &  &  &  &  &  \\
\end{tabular*}
{\rule{\temptablewidth}{1pt}}
\end{center}
\end{table}

\subsubsection{Transport properties}

Fig. 2 gives the resistivity measurement result for the BaFe$_{2}$Se$_{2}$O
single-crystalline sample. The room-temperature resistivity measured is as
high as 100 $\Omega $ cm which is comparable to those of La$_{2}$O$%
_{2} $Fe$_{2}$OSe$_{2}$\cite{ZhuJX},
Na$_{2}$Fe$_{2}$Se$_{2}$O\cite{ChenGF} and
Ba$_{2}$F$_{2}$Fe$_{2}$OSe$_{2}$\cite{BaF}. In the temperature
dependence of resistivity, an insulating behavior is observed, and
the resistivity divergence in the low temperature region below
150 K can be fitted by the model of variable-range-hopping: $\rho (T)$ = $%
\rho _{0}$exp($T_{0}$/$T$)$^{\frac{1}{4}}$. This result was also
observed in many 3d transition metal compounds in which the
correlation is assume to be strong.

\begin{figure}
\includegraphics[width=8.5cm]{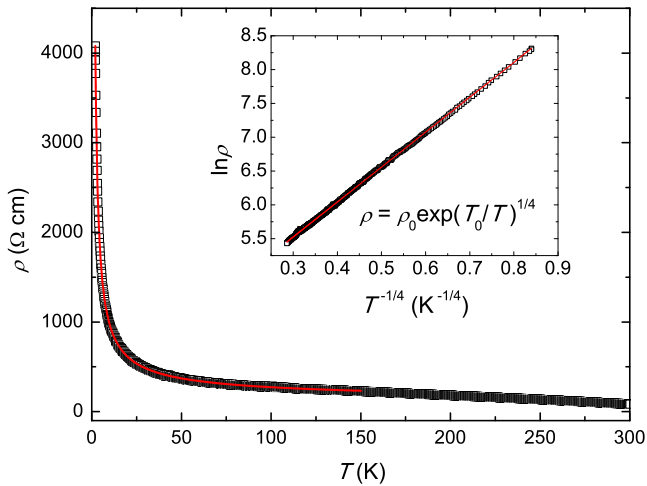}
\caption{(Color online) Temperature dependence of resistivity for the BaFe$%
_{2}$Se$_{2}$O single-crystalline sample. The inset shows that ln$\protect%
\rho$ is directly proportional to $T^{-1/4}$ below 150 K.}
\label{fig2}
\end{figure}

The temperature dependence of dc magnetic susceptibility for the BaFe$_{2}$Se%
$_{2}$O single-crystalline sample is presented in Fig. 3(a). The dc
magnetic susceptibility measurement was carried out under a magnetic
field of 20000 Oe in zero-field cooling (ZFC) process. We can see a
kink associated with the AF transition at about 240 K
in the $\chi(T)$ curve. This transition is indicated more clearly by
the peak in the d($\chi$$T$)/d$T$ versus $T$ curve, as shown in the inset
of Fig. 3(a). Similar AF transition is
observed in the parent compounds of cuprates and some iron-based superconductors%
\cite{Dagotto,122,HanF}. Above 240 K, the magnetic susceptibility
exhibits a continuing enhancement, which is observed in many
two-dimensional magnetic systems including the parent compounds of
iron-based superconductors\cite{122,HanF}
and is explained as due to the short-range correlation of the local moments\cite%
{ZhangGM}. The continuing enhancement leads to a broad maximum around $T_{max}$ =
450 K. The ratio of $T_N$/$T_{max}$ is usually used to estimate the extent
of the low-dimensional magnetic correlations. As $T_N$/$T_{max}$ in our BaFe$%
_{2}$Se$_{2}$O is 0.53, it can be concluded that BaFe$_{2}$Se$_{2}$O is a
quasi-two-dimensinal magnetic system.

\begin{figure}
\includegraphics[width=8.5cm]{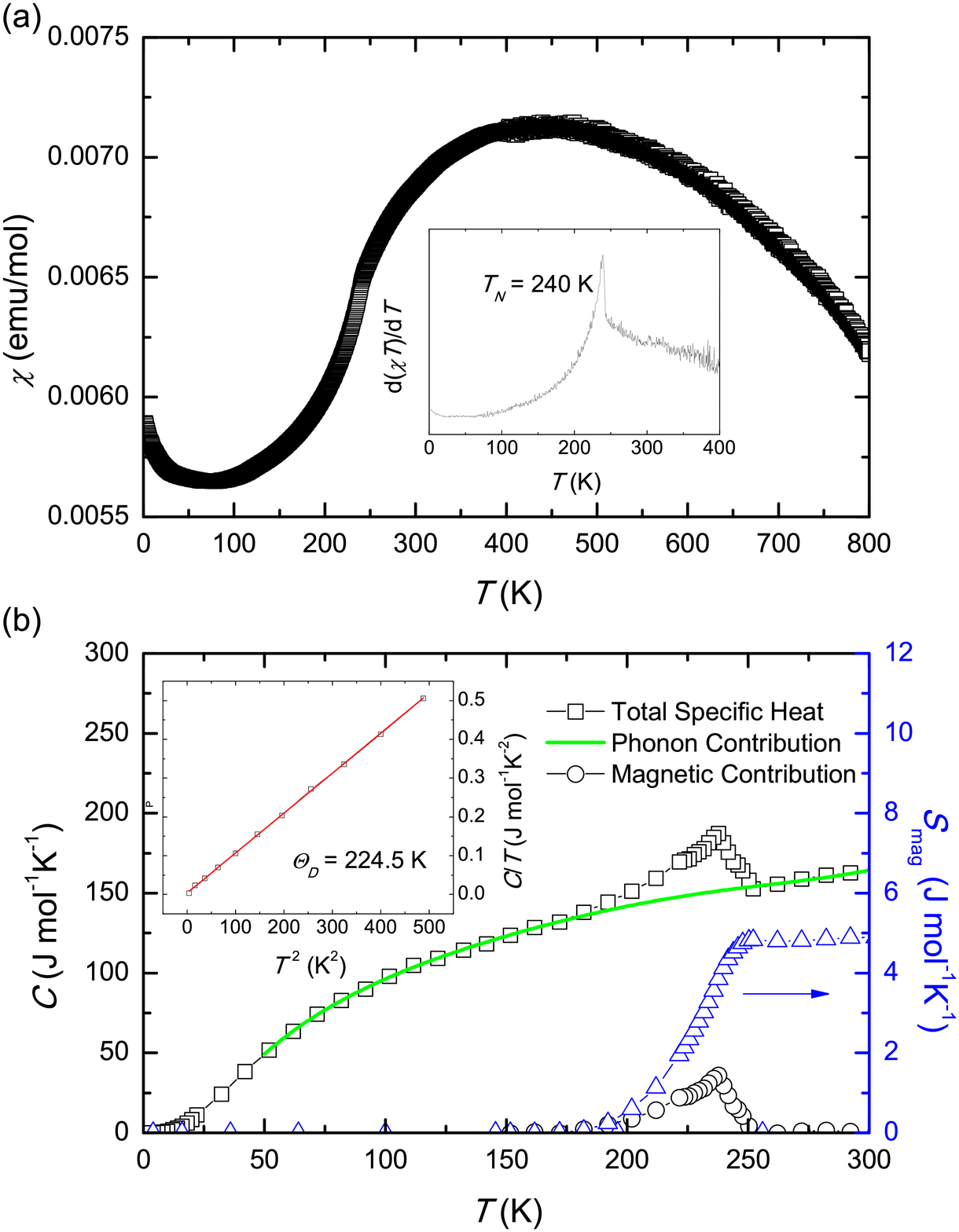}
\caption{(Color online) (a) Temperature dependence of dc magnetic
susceptibility for the BaFe$_{2}$Se$_{2}$O single-crystalline sample in
zero-field-cooling (ZFC) process at a magnetic field of H = 20000 Oe. The
inset shows the d($\chi$$T$)/d$T$ versus $T$ curve. We can see a peak
associated with an AF transition at 240 K. (b) Temperature
dependence of specific heat for the BaFe$_{2}$Se$_{2}$O single-crystalline
sample. The green solid curve represents the phonon contribution, fitted by
a polynomial. After deducting the phonon contribution, the specific heat
contributed by magnetic shows a peak at 240 K. Part of the magnetic entropy has been dissipated above 240 K.}
\label{fig3}
\end{figure}

In order to further understand the magnetic transition, we
performed a specific heat measurement, as shown in Fig. 3(b).
Provided the negligible contribution by the magnons at very low
temperatures ($T$ $\ll $ $T_{N}$), the specific heat can be
separated into electronic and phonon parts: $C(T)$ = $\gamma
$$T$ + $\beta
$$T^{3}$. We find that the low temperature data show no electronic contribution, consistent with the
insulating ground state. From the slope ($\beta $) of the straight
line in the $C/T$ vs $T^{2}$ plot (as shown in the inset of Fig.
3(b)), we obtain the Debye
temperature $\Theta _{D}$ = 224.5 K using the formula $\Theta _{D}$ = $%
[12\pi ^{4}NR/(5\beta )]^{\frac{1}{3}}$, where $N$ is the number of atoms in
the chemical formula and $R$ refers to the ideal gas constant 8.314 J mol$^{-1}$ K$%
^{-1}$. Since the Debye model fails to reproduce the $C(T)$ data at
high temperatures, we employ a polynomial to fit the phonon
contribution in the temperature range from 50 to 300 K. Assuming
that the total specific heat consists of phonon and magnetic
components, the magnetic contribution can be obtained simply by the
subtraction of the phonon contribution. A heat-capacity peak appears
at 240 K which is exactly the temperature where the magnetic
susceptibility exhibits a sharp step (as illustrated by the
derivative d($\chi$$T$)/d$T$). The magnetic entropy can be
calculated using the integral $S_{mag}$ = $\int_{0}^{T}C_{mag}/T$ d$T$. The result
gives $S_{mag}$ = 4.89 J mol$^{-1}$ K$^{-1}$. Part of the magnetic entropy has been dissipated above 240 K. Although the AF occurs at about 240 K, as reflected from the d($\chi$$T$)/d$T$ data, we find a significant contribution of entropy due to the strong AF fluctuation above 240 K.

\subsection{Band Structure Calculations}

To have a comprehensive understanding on the physical properties of BaFe$_{2}
$Se$_{2}$O, we also perform the density-functional theory calculations based
on the local density approximation (LDA) to density functional theory with
the full-potential, all-electron, linear-muffin-tin-orbital method\cite{LMTO}%
. To see the basic features of the electronic structure, we first do
nonmagnetic LDA\ calculation, and show the band structure in Fig.
4a. Although with strong hybiridization, the Fe, Se and O\ bands are
separated as shown in Fig. 4a. The six bands located from -6.6 to
-5.0 eV basically come from O 2\textit{p} states, the twelve bands
located from -5.0 to -3.4 eV are mainly contribuated by Se 4$p$
states, while the twenty Fe 3$d$ bands distribute from -1.4 to 1.2
eV. The non-coplanar honeycomb-like
structure significantly narrows the band, consequently, the bandwidth of Fe 3$%
d$, Se 4$p$ and O 2$p$ bands are considerablely narrower than these of
Fe-based superconductor and La$_{2}$O$_{2}$Fe$_{2}$OSe$_{2}$\cite{ZhuJX}.

\begin{figure}
\includegraphics[width=8.5cm]{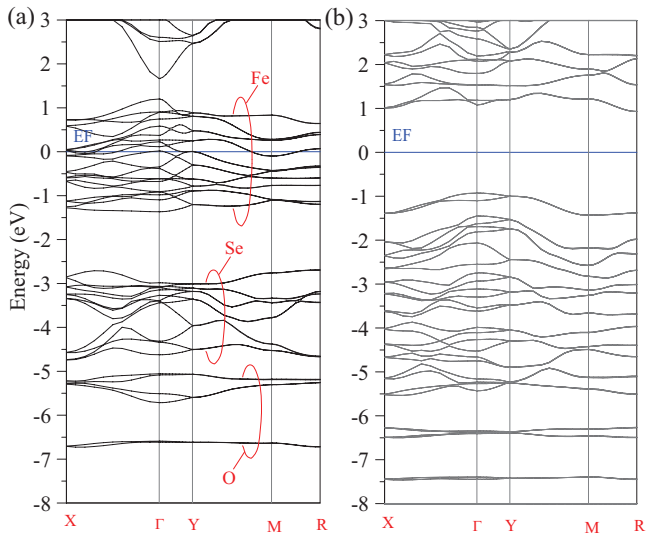}
\caption{Band structure from LDA\ calculation, and LDA+\textit{U} calculation with
\textit{U} = 6.0 eV.}
\end{figure}

We then perform the spin polarized calculation. We first evaluate the
interlayer magnetic exchange interaction based on our linear response scheme%
\cite{Linear response J}. Our numerical results show that similar
with Fe-based superconductor\cite{J-FeAs}, the exchange process
happens almost completely in the Fe-Se-O layer, and the inter-layer
exchange interaction is negligible. This result is consistent with
the observed two-dimensional magnetic behavior.\ To search the
ground state magnetic ordering configuration, we consider four spin
alignments within layer as following: FM, AFM-1 (Fe1 and Fe2, Fe3
and Fe4 couple ferromagnetically while Fe2 and Fe3 have opposite
spin orientation), AFM-2 (Fe1 and Fe2, Fe3 and Fe4 couple
antiferromagnetically while Fe2 and Fe3 have the same spin orientation), AFM-3 (Fe atom couples with all of three nearest
neighbor Fe antiferromagnetically). The three kinds of AFM spin alignments are plotted in Fig. 5. The LSDA\ calculation predicts
that the AFM-3 has the lowest total energy, and the magnetic moment
mainly locates at Fe site. It is interesting to notice that the
calculated magnetic moment (about 3.0 $\mu _{B}$)\ is not sensitive
to the magnetic configuration. While for iron arsenides, the
magnetism is quite itinerant and the magnetic moment strongly
dependent on the magnetic configuration. Although can explain the
AFM\ behavior, LSDA calculation fails to reproduce the observed
insulating behavior.

\begin{figure}
\includegraphics[width=8.5cm]{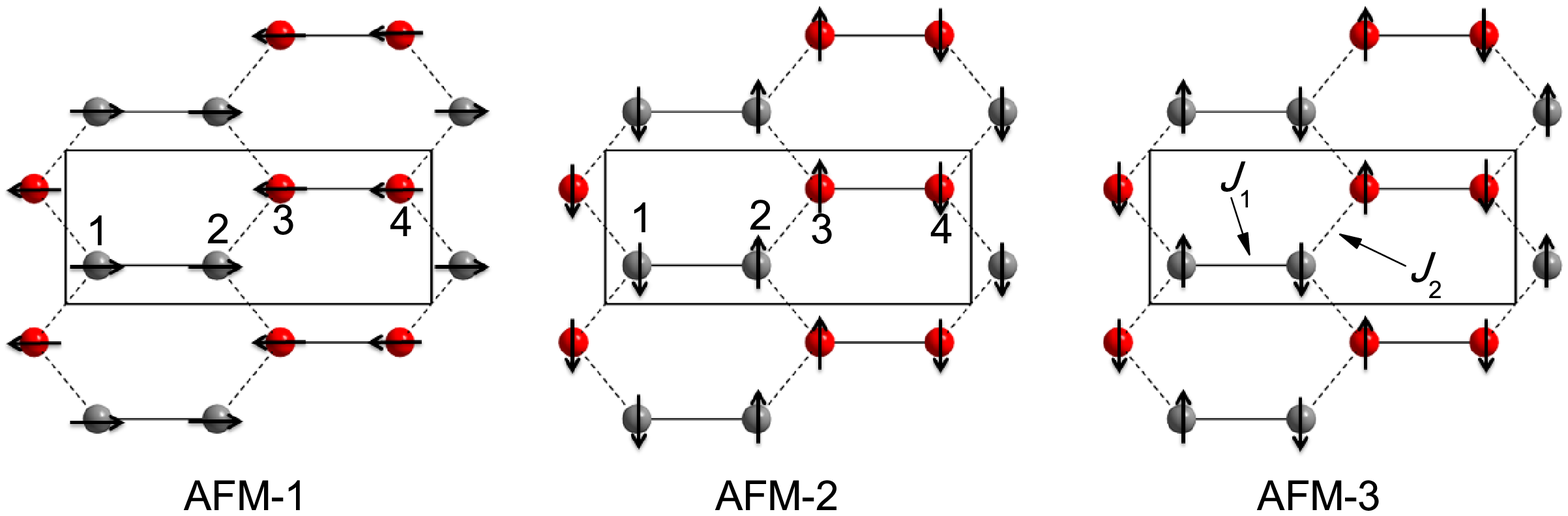}
\caption{Three kinds of possible AFM spin alignments within layer. AFM-3 has the lowest total energy.}
\end{figure}

It is well known that the Coulomb interaction for 3\textit{d}
orbital is significant, and estimates for the values of \textit{U}
have been recently done yielding \textit{U} $\approx$ 5.0 eV in
Fe-based superconductors\cite{GW-U}. Although the accurate value of
\textit{U} is not known for BaFe$_{2}$Se$_{2}$O, with narrower bands
we generally expect it is larger than that of Fe-based
superconductor. We therefore utilize LDA+\textit{U} scheme\cite{LDA+U}, which
is adequate for the magnetically ordered insulating ground
states\cite{DMFT}, and set \textit{U} = 6.0 eV. Same with LSDA calculation,
LDA+\textit{U} calculation also predicts AFM-3 as the ground state, and the
obtained magnetic moment is not depending on the magnetic
configuration. As shown in Fig.4b, electronic correlation
significantly changes the band structure and results in a
considerable effective gap (about 1.86 eV). Our linear response calculation\cite%
{J-FeAs} shows that the magnetic exchange interaction is short range
and the magnetic coupling further than the second nearest neighbor
is almost equal to zero. The magnetic exchange interactions between
all of three nearest neighbor Fe-Fe are AFM sign, which is
consistent with the total energy comparison. The strength of
exchange coupling ($J_1$ = 96 meV and $J_2$ = 31 meV)\ is similar with
that of LaFeAsO\cite{J-FeAs,J-1111}, which may account for the fact
of similar magnetic susceptibility above the Neel temperature. We
also vary parameter \textit{U} between 2 and 8 eV for Fe 3\textit{d}
electrons to see what effects the on-site Coulomb repulsion would
bring to the electronic structure of BaFe$_{2}$Se$_{2}$O. Our
numerical calculation confirm that \textit{U} does not change the qualitative
results: AFM-3 is the ground state, the magnetism is local in
nature, and the Coulomb interaction is essential for opening the
band gap. The real magnetic structure will be determined by neutron diffraction method, just like the case of $Re_2$O$_3$Fe$_2$Se$_2$\cite{Evans,NiN}.

\subsection{Concluding Remarks}

In summary, a iron-based compound BaFe$_2$Se$_2$O was synthesized
successfully. It has an orthorhombic structure with the space group
of $Pmmn$. Strictly speaking, there is not a  square-planar Fe
lattice in it, instead, it contains a honeycomb-like Fe sub-lattice.
The model of variable-range-hopping is used to explain the
resistivity divergence in the low temperature region. An AF ordering
transition at 240 K is observed both from the dc magnetic
susceptibility and specific heat. Band structure calculation reveals the narrowing of Fe 3$d$ bands near the Fermi energy, which leads to the
localization of magnetism and the Mott insulating behavior. The distances between the Fe atoms (3.1382(13) {\AA} and 3.2260(12) {\AA}) perhaps are responsible for the characters, since they are so large that the overlap between the 3$d$-orbitals becomes very small. Linear
response calculation further indicates a strong in-plane AF exchange
$J$, this can account for the enhanced magnetic susceptibility
(which has a maximum at about 450 K) above the Neel temperature.
Considering the fact that the Mottness and AF order are commonly
occurring in cuprates and the iron pnictides, it would be
interesting to know whether it is possible to induce
superconductivity by either chemical doping or applying high
pressure in our present system. This investigation is underway.

\subsection{Acknowledgments}

We are grateful to Qimiao Si and Igor I. Mazin for helpful discussions. This work is supported by the Natural Science Foundation of China,
the Ministry of Science and Technology of China (973 project:
2011CBA00102) and PAPD.

\end{document}